\documentclass[showpacs,preprintnumbers,twocolumn,amsmath,amssymb]{revtex4-1}
\usepackage{graphicx}
\usepackage{dcolumn}
\usepackage{bm}
\usepackage{epsfig}

\newcommand{\dhd}{{\textstyle d}
\lower.03ex\hbox{\kern-0.40em$^{\scriptstyle-}$}\kern-0.08em{}}  

\newcommand{\half}{{1\over 2}}
\newcommand{\bu}{{\bullet}}

\newcommand{\calr}{{\cal R}}

\newcommand{\calz}{{\cal Z}}

\newcommand{\halu}{\hat{\cal U}} 
\newcommand{\hsi}{\hat{\psi}} 
\begin{document}

\preprint{JLAB-THY-10-1231}
\preprint{LPT-ORSAY 10-74}
\preprint{CPHT-RR081.0910}

\title{Photon impact factor in the next-to-leading order}

\author{Ian Balitsky }
\affiliation{
Physics Dept., ODU, Norfolk VA 23529,  and\\
Theory Group, Jlab, 12000 Jefferson Ave, Newport News, VA 23606}
\email{balitsky@jlab.org}

\author{
Giovanni A. Chirilli}
\address{Centre de Physique ThŽorique,
Ecole polytechnique, CNRS,
91128 Palaiseau, France and\\
LPT, Universit\'e Paris-Sud, CNRS, 91405 Orsay, France\\
E-mail: chirilli@cpht.polytechnique.fr}

\date{\today}

\begin{abstract}
An analytic coordinate-space expression for the next-to-leading order photon impact factor for small-$x$ deep inelastic scattering
is calculated using the operator expansion in Wilson lines.

\end{abstract}



\maketitle

\section{\label{sec:in}Introduction }
It is well known that the small-$x$ behavior of structure functions of deep inelastic scattering is determined by the hard pomeron contribution. 
In the leading order the pomeron intercept is determined by the BFKL equation \cite{bfkl}  and the pomeron residue (the $\gamma^\ast\gamma^\ast$-pomeron vertex) 
is given by the so-called impact factor. To find the small-$x$ structure functions in the next-to-leading order, one needs to know both the pomeron intercept
and the impact factor. The NLO pomeron intercept was found many years  ago \cite{nlobfkl} but the analytic expression for the NLO impact factor is obtained
for the first time in the present paper.

 We calculate the NLO impact factor  using the high-energy operator expansion of T-product of two vector currents in Wilson lines (see e.g the reviews 
 \cite{mobzor,nlolecture}). Let us recall the general logic 
 of an operator expansion. In order to find a certain asymptotical behavior of an amplitude by OPE one should
  \begin{itemize}
\item Identify the relevant operators and  factorize an amplitude into a product of coefficient functions and matrix elements of these operators
\item Find the evolution equations of the operators with respect to the factorization scale
\item Solve these evolution equations
\item Convolute the solution  with the initial conditions for the evolution and get the amplitude.
\end{itemize}
Since we are interested in the small-$x$ asymptotics of DIS it is natural to factorize in rapidity: we introduce the rapidity divide $\eta$ which separates the ``fast'' gluons 
from the ``slow'' ones. 

 As a first step, we integrate 
over gluons with rapidities $Y>\eta$ and leave the integration over $Y<\eta$ for later time, see Fig. 2. 
It is convenient to use the background field formalism: we integrate over gluons with $\alpha>\sigma=e^\eta$ and leave gluons with $\alpha<\sigma$ as a background field, to
be integrated over later.  Since the rapidities of the background
gluons are very different from the rapidities of gluons in our Feynman diagrams, the background field can be taken in the form of a shock wave due to the Lorentz contraction.
To derive the expression of a quark (or gluon) propagator in this shock-wave background we represent the propagator as a path integral over various trajectories,
each of them weighed with the gauge factor Pexp$(ig\int\! dx_\mu A^\mu)$ ordered along the propagation path. Now, since the shock wave is very thin, quarks (or gluons) do not
have time to deviate in transverse direction so their trajectory inside the shock wave can be approximated by a segment of the straight line. Moreover, since there is no external field 
outside the shock wave, the integral over the segment of straight line can be formally extended to $\pm\infty$ limits yielding the Wilson-line 
gauge factor
\begin{eqnarray}
&&\hspace{-0mm} 
 U^\eta_x~=~{\rm Pexp}\Big[ig\!\int_{-\infty}^\infty\!\! du ~p_1^\mu A^\sigma_\mu(up_1+x_\perp)\Big],
 \nonumber\\
 &&\hspace{-0mm} 
A^\eta_\mu(x)~=~\int\!d^4 k ~\theta(e^\eta-|\alpha_k|)e^{ik\cdot x} A_\mu(k)
\label{cutoff}
\end{eqnarray}
where  the  Sudakov variable $\alpha_k$ is defined as usual,  $k=\alpha_kp_1+\beta_kp_2+k_\perp$.
(We define the light-like vectors $p_1$ and $p_2$ such that $q=p_1-x_Bp_2$  and $p_N=p_2+{m_N^2\over s}p_1$ where $p_N$ is the nucleon momentum).
The structure of the propagator in a shock-wave background looks as follows (see Fig. 1): \\ 
$\big[$Free propagation from initial point $x$ to the point of intersection with the shock wave $z\big]$\\
$\times$ $\big[$Interaction
with the shock wave described by the Wilson-line operator $U_z\big]$\\
$\times$ $\big[$Free propagation from point of interaction $z$ to the final point $y\big]$. \\
\begin{figure}[htb]
\includegraphics[width=50mm]{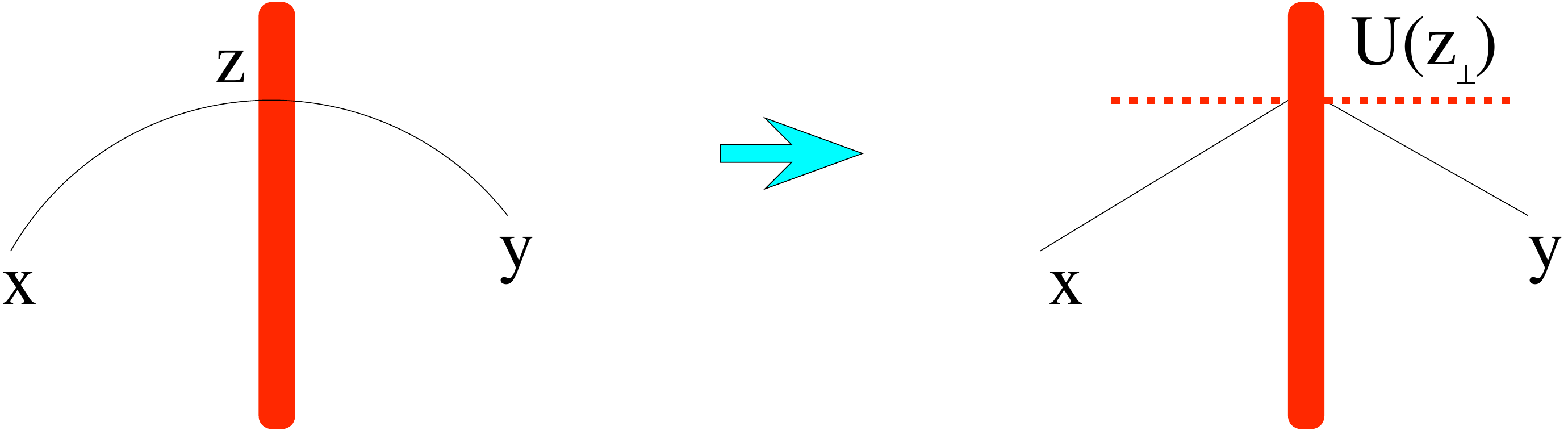}
\caption{Propagator in a shock-wave background}
\label{fig:shwaveprop}
\end{figure}

The explicit form of quark propagator in a shock-wave background can be taken from Ref. \cite{npb96}
%
\begin{eqnarray}
&&\hspace{-1mm}
\langle\hsi(x)\bar\hsi(y)\rangle~
\label{kvprop}\\
&&\hspace{-1mm}
\stackrel{x_\ast>0>y_\ast}{=}~\!\int\! d^4z~\delta(z_\ast){(\not\! x-\not\! z)\over 2\pi^2(x-z)^4}\not\! p_2U_z{(\not\! z-\not\! y)\over 2\pi^2(x-z)^4}
\nonumber
\end{eqnarray}
where we label operators by hats as usual.
Hereafter use the notations $x_\ast=p_2^\mu x_\mu={\sqrt{s}\over 2}x^+$, $x_\bullet=p_1^\mu x_\mu={\sqrt{s}\over 2}x^-$ (and our metric is (1,-1,-1,-1)).
Note that the Regge limit in the coordinate space corresponds to $x_\ast\rightarrow\infty, y_\ast\rightarrow -\infty$ while $x_\perp,y_\perp$ are fixed, see the discussion
in Refs. \cite{nlobfklconf, confamp}.

\begin{figure}[htb]
\includegraphics[width=40mm]{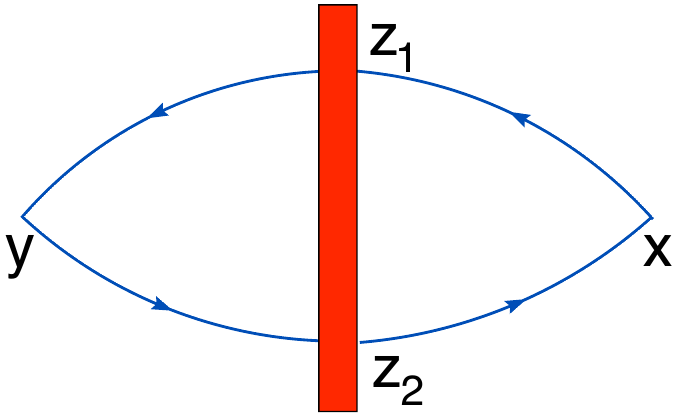}
\caption{Impact factor in the leading order. Solid lines represent quarks. \label{fig:loif}}
\end{figure}

As we mentioned above, the result of the integration over the rapidities $Y>\eta$ gives the impact factor - the amplitude of the transition of virtual photon 
in two-Wilson-lines operators (sometimes called ``color dipole'').
The LO impact factor is a product of two propagators (\ref{kvprop}), see Fig. \ref{fig:loif} 
%
\begin{eqnarray}
&&\hspace{-1mm}
\langle \bar{\hsi}(x)\gamma^\mu \hsi(x)\bar{\hsi}(y)\gamma^\nu \hsi(y) \rangle_A~ =~
\label{loif}\\
&&\hspace{-1mm}
=~{s^2\over 2^9\pi^6x_\ast^2y_\ast^2}\int d^2z_{1\perp}d^2z_{2\perp}
 {{\rm tr}\{U_{z_1}U^\dagger_{z_2}\}\over (\kappa\cdot\zeta_1)^3(\kappa\cdot\zeta_2)^3}
\nonumber\\
&&\hspace{-1mm}
\times~{\partial^2\over\partial x^\mu\partial y^\nu}
\big[2(\kappa\cdot\zeta_1)(\kappa\cdot\zeta_2)-\kappa^2(\zeta_1\cdot\zeta_2) \big]~+~O(\alpha_s)
\nonumber
\end{eqnarray}
Here we introduced the conformal vectors \cite{penecostalba,penedones} 
\begin{eqnarray}
&&\hspace{-5mm}
\kappa~=~{\sqrt{s}\over 2x_\ast}({p_1\over s}-x^2p_2+x_\perp)-{\sqrt{s}\over 2y_\ast}({p_1\over s}-y^2p_2+y_\perp)
\nonumber\\
&&\hspace{-1mm}
\zeta_i~=~\big({p_1\over s}+z_{i\perp}^2 p_2+z_{i\perp}\big),~\label{cratios}
\end{eqnarray}
and the notation $\calr~\equiv~{\kappa^2(\zeta_1\cdot\zeta_2)\over 2(\kappa\cdot\zeta_1)(\kappa\cdot\zeta_2)}$.
The above equation is explicitly M\"obius invariant. In addition, it is easy to check that ${\partial\over\partial x_\mu}$(r.h.s)=0.

Our goal is the NLO contribution to the r.h.s. of Eq. (\ref{loif}), but first let us briefly discuss the three remaining steps of 
the high-energy OPE. 
The evolution equation for color dipoles has the form \cite{npb96,yura}
\begin{eqnarray}
&&\hspace{-9mm}
{d\over d\eta}{\rm tr}\{\hat{U}^\eta_{z_1} \hat{U}^{\dagger\eta}_{z_2}\}~
=~{\alpha_s\over 2\pi^2}
\!\int\!d^2z_3~
{z_{12}^2\over z_{13}^2z_{23}^2}
~[{\rm tr}\{\hat{U}^\eta_{z_1} \hat{U}^{\dagger\eta}_{z_3}\}
\label{nlobk}\\
&&\hspace{-9mm}
\times~{\rm tr}\{\hat{U}^\eta_{z_3} \hat{U}^{\dagger\eta}_{z_2}\}
-N_c{\rm tr}\{\hat{U}^\eta_{z_1} \hat{U}^{\dagger\eta}_{z_2}\}]   ~+~{\rm NLO~contribution}
\nonumber
\end{eqnarray}
(To save space, hereafter  $z_i$ stand for $z_{i\perp}$ so $z_{12}^2\equiv z_{12\perp}^2$ etc.)
The explicit form of the NLO contributions can be found in Refs. \cite{nlobk,nlobksym,nlolecture} while the agrument
of the coupling constant in the above equation (following from the NLO calculations) is discussed in Refs. (\cite{prd75,kw1}).

The next two steps, solution of the evolution equation (\ref{nlobk}) with appropriate initial conditions and the eventual
comparison with experimental DIS data are discussed in many papers (see e.g. \cite{solutions}).

\section{Calculation of the NLO impact factor}

Now we would like to  repeat the same steps of operator expansion at the NLO accuracy. 
A general form of the expansion of  T-product of the electromagnetic currents 
 in color dipoles looks as follows:
\begin{eqnarray}
&&\hspace{-1mm}
 (x-y)^4T\{\bar{\hsi}(x)\gamma^\mu \hsi(x)\bar{\hsi}(y)\gamma^\nu \hsi(y)\}~
 \nonumber\\
&&\hspace{-1mm}=~\int\! {d^2z_1d^2z_2\over z_{12}^4}~\Big\{I_{\mu\nu}^{\rm LO}(z_1,z_2)\big[1+{\alpha_s\over\pi}\big]
 {\rm tr}\{\hat{U}^\eta_{z_1}\hat{U}^{\dagger \eta}_{z_2}\}
 \nonumber\\
&&\hspace{-1mm}
+\int\! d^2z_3~I_{\mu\nu}^{\rm NLO}(z_1,z_2,z_3;\eta)
\nonumber\\
&&\hspace{-1mm}
\times~
[ {\rm tr}\{\hat{U}^\eta_{z_1}\hat{U}^{\dagger \eta}_{z_3}\}{\rm tr}\{\hat{U}^\eta_{z_3}\hat{U}^{\dagger \eta}_{z_2}\}
 -N_c {\rm tr}\{\hat{U}^\eta_{z_1}\hat{U}^{\dagger \eta}_{z_2}\}]\Big\}
 \label{opeq}
 \end{eqnarray}
The structure of the NLO contribution is clear from the topology of diagrams in the shock-wave background, see Fig. \ref{fig:nloif} below.
Also, the term $\sim~1+{\alpha_s\over \pi}$ can be restored from the requirement that at $U=1$ (no shock wave) one should get the perturbative series for the 
polarization operator $1+{\alpha_s\over \pi}+O(\alpha_s^2)$.

In our notations
\begin{eqnarray}
&&\hspace{-1mm}
I^{\rm LO}_{\mu\nu}(z_1,z_2)~=~
 {\calr^2\over \pi^6(\kappa\cdot\zeta_1)(\kappa\cdot\zeta_2)}
\nonumber\\
&&\hspace{-1mm}
\times~{\partial^2\over\partial x^\mu\partial y^\nu}
\big[(\kappa\cdot\zeta_1)(\kappa\cdot\zeta_2)-\half\kappa^2(\zeta_1\cdot\zeta_2) \big].~
\label{loif}
\end{eqnarray}
which corresponds to the well-known expression for the LO impact factor in the momentum space.

The NLO impact factor is given by the diagrams shown in Fig. \ref{fig:nloif}. 
\begin{figure}[htb]
\includegraphics[width=70mm]{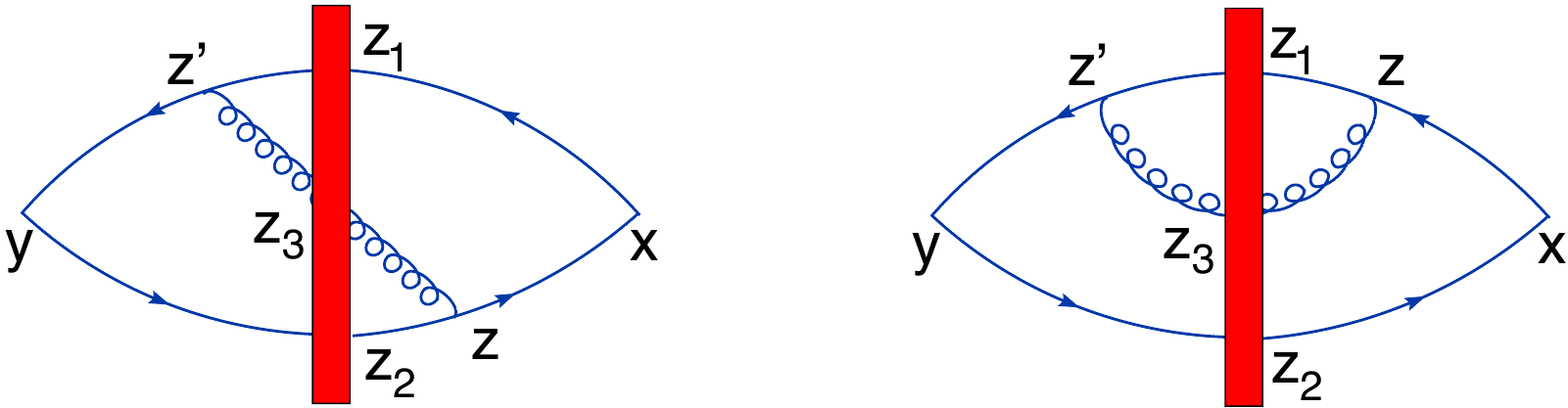}

\vspace{3mm}
\includegraphics[width=70mm]{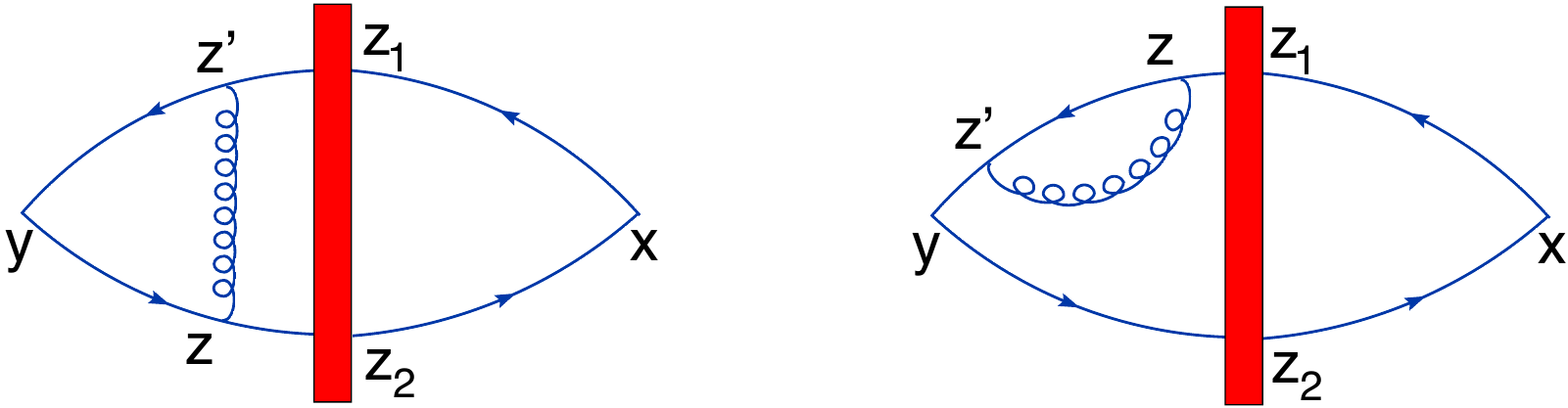}
\caption{Impact factor in the next-to-leading order. \label{fig:nloif}}
\end{figure}
The calculation of these diagrams
is similar to the calculation of the NLO impact factor for scalar currents in ${\cal N}=4$ SYM carried out in our previous paper \cite{nlobksym}.
The gluon propagator in the shock-wave background at $x_\ast>0>y_\ast$ in the light-like gauge $p_2^\mu A_\mu=0$ is given by \cite{prd99,balbel} 
\begin{eqnarray}
&&\hspace{-1mm}
\langle \hat{A}^a_\mu(x)\hat{A}^b_\nu(y)\rangle
~
\stackrel{x_\ast>0>y_\ast}{=}~-{i\over 2}\int d^4 z~\delta(z_\ast)~
\label{gluprop}\\
&&\hspace{-1mm}
\times~{x_\ast g^\perp_{\mu\xi}-p_{2\mu}(x-z)^\perp_\xi \over \pi^2[(x-z)^2+i\epsilon]^2}\;U^{ab}_{z_\perp}
{1\over\partial_\ast^{(z)}}~{y_*\delta^{\perp\xi}_\nu
 - p_{2\nu}(y-z)_\perp^\xi\over \pi^2[(z-y)^2+i\epsilon]^2}
 \nonumber
\end{eqnarray}
where ${1\over \partial_\ast}$ can be either ${1\over \partial_\ast+i\epsilon }$ or 
${1\over \partial_\ast-i\epsilon }$ which leads to the same result. (This is obvious for the leading order and 
correct in NLO after subtraction of the leading-order contribution, see Eq. (\ref{nloify}) below).

The diagrams in Fig. \ref{fig:nloif}a,b can be calculated using the conformal integral
\begin{eqnarray}
&&\hspace{-3mm}
\int\! d^4z                                                             
~{\not\!{x}-\not\!{z}\over (x-z)^4}\gamma_\mu
{\not\!{z}-\not\!{y}\over(z-y)^4}{z_\nu\over z^4}-\mu\leftrightarrow\nu~
\nonumber\\
&&\hspace{-3mm} 
=~{\pi^2\over x^2y^2(x-y)^2}\Big[-\!\not\! x\gamma_\mu\!\not\! y\Big({x_\nu\over x^2}+{y_\nu\over y^2}\Big)
\nonumber\\
&&\hspace{-3mm} 
+~\half(\!\not\! x\gamma_\mu\gamma_\nu-\gamma_\mu\gamma_\nu\!\not\! y)
+2x_\mu y_\nu{\!\not\! x-\!\not\! y\over (x-y)^2}\Big]~-~\mu\leftrightarrow\nu~~~
\label{confintegral}
\end{eqnarray}
which gives the 3-point $\psi\Bar{\psi}F_{\mu\nu}$ Green function in the leading order in $g$. 
Using Eqs. (\ref{kvprop}), (\ref{gluprop}) and (\ref{confintegral}), performing integrals over $z_\bu$'s and 
taking traces one gets after some algebra the NLO contribution of diagrams in Fig. \ref{fig:nloif} 
in the form 
\begin{equation}
I^{\rm Fig.\ref{fig:nloif}}_{\mu\nu}(z_1,z_2,z_3)~=~\tilde{I}_1^{\mu\nu}(z_1,z_2,z_3)+I_2^{\mu\nu}(z_1,z_2,z_3)
\label{fig3vklad}
\end{equation}
where
\begin{eqnarray}
&&\hspace{-1mm}
~\tilde{I}_1^{\mu\nu}(z_1,z_2,z_3)~
\label{tildei1}\\
&&\hspace{-1mm}=~{\alpha_s\over 4\pi^2}I^{\rm LO}_{\mu\nu}(z_1,z_2){(\zeta_1\cdot\zeta_2)\over (\zeta_1\cdot\zeta_3) (\zeta_1\cdot\zeta_3)}
\!\int_0^\infty\!{d\alpha\over\alpha}e^{i\alpha {s\over 4}\sigma\calz_3}
\nonumber
\end{eqnarray}
and
\begin{eqnarray}
&&\hspace{-1mm}
(I_2)_{\mu\nu}(z_1,z_2,z_3)~
=~{\alpha_s\over 16\pi^8}{\calr^2\over(\kappa\cdot\zeta_1)(\kappa\cdot\zeta_2)}
\label{nloifoton}\\
&&\hspace{-1mm}
\times~
\Bigg\{{(\kappa\cdot\zeta_2)\over(\kappa\cdot\zeta_3)}
{\partial^2\over\partial x^\mu\partial y^\nu}
\Big[-{(\kappa\cdot\zeta_1)^2\over (\zeta_1\cdot\zeta_3)}+{(\kappa\cdot\zeta_1)(\kappa\cdot\zeta_2)\over (\zeta_2\cdot\zeta_3)}
\nonumber\\
&&\hspace{-1mm}
+~{(\kappa\cdot\zeta_1)(\kappa\cdot\zeta_3)(\zeta_1\cdot\zeta_2)\over  (\zeta_1\cdot\zeta_3) (\zeta_2\cdot\zeta_3)}
-{\kappa^2(\zeta_1\cdot\zeta_2)\over (\zeta_2\cdot\zeta_3)}\Big]+{(\kappa\cdot\zeta_2)^2\over (\kappa\cdot\zeta_3)^2}
\nonumber\\
&&\hspace{-1mm}
\times~{\partial^2\over\partial x^\mu\partial y^\nu}
\Big[{(\kappa\cdot\zeta_1)(\kappa\cdot\zeta_3)\over (\zeta_2\cdot\zeta_3)}-{\kappa^2(\zeta_1\cdot\zeta_3)\over 2(\zeta_2\cdot\zeta_3)}\Big]
+(\zeta_1\leftrightarrow\zeta_2)\Bigg\}
\nonumber
\end{eqnarray}
(recall that $z_{ij\perp}^2=2(\zeta_i\cdot\zeta_j)$ and $\calz_i={4\over\sqrt{s}}(\kappa\cdot\zeta_i)$). 
We obtained this expression at $x_\ast>0>y_\ast$ but  from the conformal structure of the result it is clear 
that this expression holds true at $x_\ast<0<y_\ast$ as well.

The integral over $\alpha$ in the r.h.s. of Eq. (\ref{tildei1}) diverges. This divergence reflects the fact that the 
contributions of the diagrams in Fig. \ref{fig:nloif} is not exactly the NLO impact factor since we must subtract the 
matrix element of the leading-order contribution. Indeed,  the NLO impact factor is a coefficient function defined according to Eq. (\ref{opeq}).
To find the NLO impact factor, we consider the operator equation (\ref{opeq}) in 
the shock-wave background (in the leading order $\langle\hat{U}_{z_3}\rangle_A=U_{z_3}$):
\begin{eqnarray}
&&\hspace{-1mm}
 \langle T\{\bar{\hsi}(x)\gamma^\mu \hsi(x)\bar{\hsi}(y)\gamma^\nu \hsi(y)\}\rangle_A~
  \nonumber\\
&&\hspace{-1mm}
-\int\! {d^2z_1d^2z_2\over z_{12}^4}~I_{\mu\nu}^{\rm LO}(x,y;z_1,z_2)
 \langle{\rm tr}\{\hat{U}^\eta_{z_1}\hat{U}^{\dagger \eta}_{z_2}\}\rangle_A
 \nonumber\\
&&\hspace{-1mm}
=~\int\! {d^2z_1d^2z_2\over z_{12}^4}d^2z_3~I_{\mu\nu}^{\rm NLO}(x,y;z_1,z_2,z_3;\eta)
 \nonumber\\
&&\hspace{-1mm}
[ {\rm tr}\{U_{z_1}U^\dagger_{z_3}\}{\rm tr}\{U_{z_3}U^\dagger_{z_2}\}
 -N_c {\rm tr}\{U_{z_1}U^\dagger_{z_2}\}]
 \label{opeq1}
 \end{eqnarray}
The NLO matrix element $ \langle T\{\bar{\hsi}(x)\gamma^\mu \hsi(x)\bar{\hsi}(y)\gamma^\nu \hsi(y)\}\rangle_A$ is given by Eq. (\ref{fig3vklad}) 
while the subtracted term is
\begin{eqnarray}
&&\hspace{-1mm}
{\alpha_s\over 2\pi^2}\!\int\! {d^2z_1d^2z_2\over z_{12}^4}~I_{\mu\nu}^{\rm LO}(z_1,z_2)\!\int_0^\sigma\!{d\alpha\over\alpha}\!\int\!d^2z_3~
{z_{12}^2\over z_{13}^2z_{23}^2}
\nonumber\\
&&\hspace{-2mm}
\times~
~[{\rm tr}\{U_{z_1} U^{\dagger}_{z_3}\}{\rm tr}\{ U_{z_3} U^{\dagger}_{z_2}\}
-~N_c{\rm tr}\{U_{z_1} U^{\dagger}_{z_2}\}]   
\label{subtracterm}
\end{eqnarray}
as follows from Eq. (\ref{nlobk}).
The $\alpha$ integration is cut from above by $\sigma=e^\eta$ in accordance with the definition of operators $\hat{U}^\eta$, see Eq.  (\ref{cutoff}). 
Subtracting  (\ref{subtracterm}) from  Eq. (\ref{fig3vklad}) we get
\begin{eqnarray}
&&\hspace{-3mm}
I^{\rm NLO}_{\mu\nu}(z_1,z_2,z_3;\eta)~=~I_1^{\mu\nu}(z_1,z_2,z_3;\eta)+I_2^{\mu\nu}(z_1,z_2,z_3),
\nonumber\\
&&\hspace{-3mm}
I_1^{\mu\nu}(x,y;z_1,z_2,z_3;\eta)~
\nonumber\\
&&\hspace{-3mm}
=~{\alpha_s\over 2\pi^2}I^{\rm LO}_{\mu\nu}(z_1,z_2){z_{12}^2\over z_{13}^2z_{23}^2}
\Big[\!\int_0^\infty\!{d\alpha\over\alpha}~e^{i\alpha {s\over 4}\calz_3}   
-\!\int_0^\sigma\!{d\alpha\over\alpha}\Big]
\nonumber\\
&&\hspace{-2mm}
=~-{\alpha_s\over 2\pi^2}I^{\rm LO}_{\mu\nu}
{z_{12}^2\over z_{13}^2z_{23}^2}
\big[\ln{\sigma s\over 4}\calz_3-{i\pi\over 2}+C\big]
 \label{nloify}
 \end{eqnarray}
Note that one should expect the NLO impact factor to be conformally invariant since it is determined by tree diagrams in Fig. \ref{fig:nloif}.
However, as discussed in Refs. \cite{nlobk,confamp,nlolecture}, formally the light-like Wilson lines are conformally (M\"obius) invariant but the 
longitudinal cutoff $\alpha<\sigma$ in Eq. (1) violates this property so the term $\sim\ln\sigma\calz_3$ in the r.h.s. of Eq. (\ref{nloify}) is not invariant. 
As was demonstrated in these papers, one can define a composite operator
in the form 
\begin{eqnarray}
&&\hspace{-2mm}
[{\rm tr}\{\hat{U}_{z_1}\hat{U}^{\dagger}_{z_2}\}\big]_a~
\label{confodipola}\\
&&\hspace{-2mm}
=~{\rm tr}\{\hat{U}^\sigma_{z_1}\hat{U}^{\dagger\sigma}_{z_2}\}
+~{\alpha_s\over 2\pi^2}\!\int\! d^2 z_3~{z_{12}^2\over z_{13}^2z_{23}^2}
[ {\rm tr}\{\hat{U}^\sigma_{z_1}\hat{U}^{\dagger\sigma}_{z_3}\}
\nonumber\\
&&\hspace{-2mm}
\times~{\rm tr}\{\hat{U}^\sigma_{z_3}\hat{U}^{\dagger\sigma}_{z_2}\}
-N_c {\rm tr}\{\hat{U}^\sigma_{z_1}\hat{U}^{\dagger\sigma}_{z_2}\}]
\ln {4az_{12}^2\over \sigma^2 sz_{13}^2z_{23}^2}~+~O(\alpha_s^2)
\nonumber
\end{eqnarray}
where $a$ is an arbitrary constant. It is convenient to choose the rapidity-dependent constant 
$a\rightarrow ae^{-2\eta}$ so that the 
$[{\rm tr}\{\hat{U}^\sigma_{z_1}\hat{U}^{\dagger\sigma}_{z_2}\}\big]_a^{\rm conf}$ 
does not depend on $\eta=\ln\sigma$ and all the rapidity dependence is encoded into $a$-dependence. Indeed,
it is easy to see that ${d\over d\eta}[{\rm tr}\{\hat{U}_{z_1}\hat{U}^{\dagger}_{z_2}\}\big]_a^{\rm conf}~=~0$ 
and ${d\over da}[{\rm tr}\{\hat{U}_{z_1}\hat{U}^{\dagger}_{z_2}\}\big]_a^{\rm conf}$ is determined by the
NLO BK kernel which is a sum of the conformal part and the running-coupling part
with our $O(\alpha_s^2)$ accuracy\cite{nlobksym,nlolecture}. 

Rewritten in terms of composite  dipoles (\ref{confodipola}), the operator expansion (\ref{opeq})  takes the form:
\begin{eqnarray}
&&\hspace{-1mm}
 T\{\bar{\hsi}(x)\gamma^\mu \hsi(x)\bar{\hsi}(y)\gamma^\nu \hsi(y)\}~
\nonumber\\
&&\hspace{-1mm}=~\int\! {d^2z_1d^2z_2\over z_{12}^4}~\Big\{I_{\mu\nu}^{\rm LO}(z_1,z_2)\big[1+{\alpha_s\over\pi}\big]
[ {\rm tr}\{\hat{U}_{z_1}\hat{U}^{\dagger }_{z_2}\}]_a
 \nonumber\\
&&\hspace{-1mm}
+\int\! d^2z_3~I_{\mu\nu}^{\rm NLO}(z_1,z_2,z_3;a)
\nonumber\\
&&\hspace{-1mm}
\times~
[ {\rm tr}\{\hat{U}_{z_1}\hat{U}^{\dagger}_{z_3}\}{\rm tr}\{\hat{U}_{z_3}\hat{U}^{\dagger}_{z_2}\}
 -N_c {\rm tr}\{\hat{U}_{z_1}\hat{U}^{\dagger}_{z_2}\}]_a\Big\}
 \label{opeconf}
 \end{eqnarray}
We need to choose the  ``new rapidity cutoff'' $a$ in such a way that all the energy dependence is included in the matrix element(s) of 
Wilson-line operators so the impact factor should not depend on energy. A suitable
choice of $a$ is given by $a_0=-\kappa^{-2}+i\epsilon=-{4x_\ast y_\ast\over s(x-y)^2}+i\epsilon$ so we obtain
\begin{eqnarray}
&&\hspace{-1mm}
 (x-y)^4T\{\bar{\hsi}(x)\gamma^\mu \hsi(x)\bar{\hsi}(y)\gamma^\nu \hsi(y)\}~
 \label{opeconfin}\\
&&\hspace{-1mm}=~\int\! {d^2z_1d^2z_2\over z_{12}^4}~\Big\{I^{\mu\nu}_{\rm LO}(z_1,z_2)\big[1+{\alpha_s\over \pi}\big]
[ {\rm tr}\{\hat{U}_{z_1}\hat{U}^{\dagger }_{z_2}\}]_{a_0}
 \nonumber\\
&&\hspace{-1mm}
+\int\! d^2z_3\Big[ {\alpha_s\over 4\pi^2}{z_{12}^2\over z_{13}^2z_{23}^2}
\Big(\ln{\kappa^2(\zeta_1\cdot\zeta_3)(\zeta_1\cdot\zeta_3)\over 2(\kappa\cdot\zeta_3)^2(\zeta_1\cdot\zeta_2)}
-2C\Big)I^{\mu\nu}_{\rm LO}
\nonumber\\
&&\hspace{-1mm}
+~I_2^{\mu\nu}\Big]
[ {\rm tr}\{\hat{U}_{z_1}\hat{U}^{\dagger}_{z_3}\}{\rm tr}\{\hat{U}_{z_3}\hat{U}^{\dagger}_{z_2}\}
 -N_c {\rm tr}\{\hat{U}_{z_1}\hat{U}^{\dagger}_{z_2}\}]_{a_0}\Big\}
 \nonumber
 \end{eqnarray}
Here the composite dipole $[{\rm tr}\{\hat{U}^\sigma_{z_1}\hat{U}^{\dagger\sigma}_{z_2}\}]_{a_0}$ is given by Eq. (\ref{confodipola}) with
$a_0=-{4 x_\ast y_\ast\over s(x-y)^2}+i\epsilon$ while  $I^{\mu\nu}_{\rm LO}(z_1,z_2)$ and $I_2^{\mu\nu}(z_1,z_2,z_3)$ are given by Eqs. (\ref{loif})
and (\ref{nloifoton}), respectively.

\section{NLO impact factor for the BFKL pomeron}

For the studies of DIS with the linear NLO BFKL equation (up to two-gluon accuracy)  we need the linearized version of Eq. (\ref{opeconfin}).
If we define 
\begin{equation}
\halu_{a}(z_1,z_2)=1-{1\over N_c}[{\rm tr}\{\hat{U}_{z_1}\hat{U}^{\dagger}_{z_2}\}]_a
\end{equation}
and consider the linearization
\begin{eqnarray}
&&\hspace{-0mm}
{1\over N_c^2} {\rm tr}\{\hat{U}_{z_1}\hat{U}^{\dagger}_{z_3}\}{\rm tr}\{\hat{U}_{z_3}\hat{U}^{\dagger}_{z_2}\}
 -{1\over N_c} {\rm tr}\{\hat{U}_{z_1}\hat{U}^{\dagger}_{z_2}\}]_{a_0}~
 \nonumber\\
&&\hspace{-0mm}
\simeq~\halu(z_1,z_2)-\halu(z_1,z_3)-\halu(z_2,z_3)
 \nonumber
 \end{eqnarray}
one of the integrals over $z_i$ in the r.h.s. of Eq. (\ref{opeconfin}) can be performed. 
The result is
\begin{eqnarray}
&&\hspace{-4mm}
{1\over N_c}(x-y)^4T\{\bar{\hsi}(x)\gamma^\mu \hsi(x)\bar{\hsi}(y)\gamma^\nu \hsi(y)\}~
\label{ifresult}\\
&&\hspace{-4mm}
=~{\partial\kappa^\alpha\over\partial x^\mu}{\partial\kappa^\beta\over\partial y^\nu}
\!\int\! {dz_1 dz_2\over z_{12}^4}~\halu_{a_0}(z_1,z_2)\big[{\cal I}_{\alpha\beta}^{\rm LO}\big(1+{\alpha_s\over\pi}\big)+ {\cal I}_{\alpha\beta}^{\rm NLO}\big]
\nonumber
\end{eqnarray}
where
\begin{equation}
{\cal I}^{\alpha\beta}_{\rm LO}(x,y;z_1,z_2)~=~ \calr^2{g^{\alpha\beta}(\zeta_1\cdot\zeta_2)-\zeta_1^\alpha\zeta_2^\beta-\zeta_2^\alpha\zeta_1^\beta\over \pi^6(\kappa\cdot\zeta_1)(\kappa\cdot\zeta_2)}
\label{loif1}
\end{equation}
(see Eq. (\ref{loif}) and 
\begin{eqnarray}
&&\hspace{-1mm}
{\cal I}^{\alpha\beta}_{\rm NLO}(x,y;z_1,z_2)~=~
{\alpha_sN_c\over 4\pi^7}\calr^2
\Bigg\{
{\zeta_1^\alpha\zeta_2^\beta+\zeta_1\leftrightarrow \zeta_2\over (\kappa\cdot\zeta_1)(\kappa\cdot\zeta_2)}
\nonumber\\
&&\hspace{-1mm}
\times~
\Big[4{\rm Li}_2(1-{\cal R})-{2\pi^2\over 3}+{2\ln {\cal R}\over 1-{\cal R}}+{\ln {\cal R}\over {\cal R}}-4\ln {\cal R} +{1\over 2{\cal R}}
\nonumber\\
&&\hspace{-1mm}
-~2+2(\ln {1\over  {\cal R}}+{1\over {\cal R}}-2)\big(\ln {1\over {\cal R}}+2C\big)-4C-{2C\over {\cal R}}\Big]
\nonumber\\
&&\hspace{-1mm}
+\Big({\zeta_1^\alpha\zeta_1^\beta\over(\kappa\cdot\zeta_1)^2}
+\zeta_1\leftrightarrow\zeta_2\Big)
\Big[{\ln {\cal R}\over {\cal R}}-{2C\over {\cal R}}+2{\ln {\cal R}\over 1-{\cal R}}-{1\over 2{\cal R}}\Big]
\nonumber\\
&&\hspace{-1mm}
+~\Big[{\zeta_1^\alpha\kappa^\beta+\zeta_1^\beta\kappa^\alpha\over (\kappa\cdot\zeta_1)\kappa^2}
+\zeta_1\leftrightarrow \zeta_2\Big]
\Big[-2{\ln {\cal R}\over 1-{\cal R}}-{\ln {\cal R}\over {\cal R}}
\nonumber\\
&&\hspace{-1mm}
+\ln {\cal R}-{3\over 2{\cal R}}+{5\over 2}+2C+{2C\over {\cal R}}\Big] 
-{2\over\kappa^2}\Big(g^{\alpha\beta}-2{\kappa^\alpha\kappa^\beta\over\kappa^2}\Big)
\nonumber\\
&&\hspace{-1mm}
+~{g^{\alpha\beta}(\zeta_1\cdot\zeta_2)\over(\kappa\cdot\zeta_1)(\kappa\cdot\zeta_2)}\Big[{2\pi^2\over 3}-4{\rm Li}_2(1-{\cal R})
-2\big(\ln {1\over  {\cal R}}+{1\over {\cal R}}
\nonumber\\
&&\hspace{-1mm}
+~{1\over 2{\cal R}^2}-3\big)\big(\ln {1\over {\cal R}}+2C\big)
+6\ln {\cal R}-{2\over  {\cal R}}+2 +{3\over 2{\cal R}^2}\Big]\Bigg\}
\nonumber\\
\label{nloifresult}
\end{eqnarray}
Here one easily recognizes five conformal tensor structures discussed in Ref. \cite{penecostalba2}.

While it is easy to see that 
\begin{eqnarray}
&&\hspace{-1mm}
{d\over dx_\mu}{1\over (x-y)^4}{\partial\kappa^\alpha\over\partial x^\mu}{\partial\kappa^\beta\over\partial y^\nu}
{\cal I}_{\alpha\beta}^{\rm LO}(x,y;z_i)~=~0
\end{eqnarray}
one should be careful when checking the electromagnetic gauge invariance in the next-to-leading order. The reason is that
the  composite dipole $\halu^{a_0}(z_1,z_2)$ depends on $x$ via the rapidity cutoff $a_0=-{4x_\ast y_\ast\over s(x-y)^2}$ so 
from Eq. (\ref{ifresult}) we get
\begin{eqnarray}
&&\hspace{-4mm}
{d\over dx_\mu}{1\over (x-y)^4}{\partial\kappa^\alpha\over\partial x^\mu}{\partial\kappa^\beta\over\partial y^\nu}
\!\int\! {dz_1 dz_2\over z_{12}^4}~\halu_{a_0}(z_1,z_2){\cal I}_{\alpha\beta}^{\rm NLO}(x,y;z_i)
\nonumber\\
&&\hspace{-4mm}
=~\Big(2{(x-y)^\mu\over (x-y)^2}-{p_2^\mu\over x_\ast}\Big){1\over (x-y)^4}
\nonumber\\
&&\hspace{-4mm}
\times~{\partial\kappa^\alpha\over\partial x^\mu}{\partial\kappa^\beta\over\partial y^\nu}\!\int\! {dz_1 dz_2\over z_{12}^4}
{\cal I}^{\rm LO}_{\alpha\beta}(x,y;z_i)
\left.a{d\over da}\halu_a(z_1,z_2)\right|_{a_0}
\end{eqnarray}
Using the leading-order BFKL equation in the dipole form (linearization of Eq. (\ref{nlobk}))
\begin{eqnarray}
&&\hspace{-1mm}
a{d\over da}\hat{\cal U}_a(z_1,z_2)
~=~{\alpha_sN_c\over 4\pi^2}\!\int\!d^2z_3{z_{12}^2\over z_{13}^2z_{23}^2}
\nonumber\\
&&\hspace{-1mm}
\times~
\big[\hat{\cal U}_a(z_1,z_3)+\hat{\cal U}_a(z_2,z_3)-\hat{\cal U}_a(z_1,z_2)\big]
\label{nlolin}
\end{eqnarray}
we obtain the following consequence of gauge invariance
\begin{eqnarray}
&&\hspace{-1mm}
{d\over dx_\mu}{1\over (x-y)^4}{\partial\kappa^\alpha\over\partial x^\mu}{\partial\kappa^\beta\over\partial y^\nu}
{\cal I}_{\alpha\beta}^{\rm NLO}(x,y;z_i)
\label{checkgaugeinv}\\
&&\hspace{-1mm}
=~{\alpha_s\over \pi^8}
{ y_\ast\over x_\ast (x-y)^6}\calr^3
\Big[\big({1\over 2\calr}-3-\ln \calr\big){\partial\ln\kappa^2\over \partial y^\nu}
\nonumber\\
&&\hspace{-1mm}
+\Big({\ln \calr\over\calr}+{5\over 2\calr}-{1\over 2\calr^2}\Big){\partial\over \partial y^\nu}[\ln(\kappa\cdot\zeta_1)+\ln(\kappa\cdot\zeta_2)]\Big]
\nonumber
\end{eqnarray}
We have verified that the expression (\ref{nloifresult}) satisfies the above equation.  

\section{Conclusions and outlook}
We have calculated the NLO impact factor for the virtual photons both in the non-linear form (\ref{opeconfin}) and 
with the  linear (two-gluon) accuracy (\ref{ifresult}).  Our results are obtained in the coordinate representation so 
the next step should be the Fourier transformation of Eq. (\ref{nloifresult}) 
which would give the momentum-space impact factor convenient for phenomenological applications
(and available at present only as a combination of numerical and analytical expressions\cite{bart1}).
The study
is in progress.

\section*{Acknowledgements}
The authors are grateful to L.N. Lipatov for valuable discussions.
This work was supported by contract
 DE-AC05-06OR23177 under which the Jefferson Science Associates, LLC operate the Thomas Jefferson National Accelerator Facility. 
 The work of G.A.C is supported by the grant ANR-06-JCJC-0084.
 
\end{document}